\documentclass[a4paper,amsmath,amssymb,superscriptaddress,aps,prb,reprint,showpacs]{revtex4-2} 
\usepackage[caption=false]{subfig}
\usepackage{xcolor,graphicx} 
\usepackage{MnSymbol} 
\usepackage{mathtools}
\usepackage{braket} 
\usepackage{tikz} 

\usepackage[utf8]{inputenc} 
\usepackage{hyperref}
\usepackage{here}
\usepackage{bbold}
\usepackage[version=4]{mhchem}

\begin{document}

\title{Robustness of the flux-free sector of the Kitaev honeycomb against environment}

\author{Alexander Sattler}
\author{Maria Daghofer}
\affiliation{
Institut f\"ur Funktionelle Materie und Quantentechnologien,
Universit\"at Stuttgart,
70569 Stuttgart,
Germany}

\date{\today}

\begin{abstract}
  The Kitaev honeycomb model (KHM) consists of spin-$1/2$
  particles on a honeycomb lattice with direction-dependent Ising-like
  interactions. It can alternatively be described in terms of non-interacting Majorana
  fermions, can be solved exactly, and has a quantum spin-liquid ground state. Open boundaries then host Majorana zero
  modes (MZMs) that are robust against some types of disorder. We
  analyze the fate of the MZMs when they couple to an environment
  via a Lindblad master equation. 
By computing the time evolution of the density matrix, we find that when decoherence occurs, the steady state is mostly the maximally mixed state. 
   Among the few exceptions is a
  parameter regime that realizes the superconducting Kitaev chain model with
  periodic boundary conditions.
    We consistently observe a
  quantum Zeno effect in the density matrix as well as in the  entropy and fidelity, while it is not found in  the energy gap of some gapped
  spin liquids.  
  We thus present a comprehensive overview over MZMs coupled to a spin
  bath that is 
  relevant to proposals to detect MZMs of Kitaev layers on surfaces
  using scanning tunneling microscopy (STM). 
\end{abstract}

\maketitle

\section{Introduction} \label{sec:introduction}

Topological order~\cite{RevModPhys.89.040502,RevModPhys.89.041004,RevModPhys.88.035005,RevModPhys.83.1057,RevModPhys.82.3045} has attracted substantial interest owing to its ability to exhibit exotic phenomena such as fractionalized excitations, anyonic quantum statistics, gapless edge modes, and ground-state degeneracies. Transitions from topologically ordered phases to topologically trivial ones are characterized by closing the bulk energy gap. Notably, the topological characteristics of these phases are intrinsically robust against certain types of disorder.

The stability of topological order and its hallmark features, such as edge states, is well established in closed quantum systems. However, because all physical systems are subject to some degree of environmental coupling, it becomes essential to understand how interactions with the environment affect these phases. In particular, the role of decoherence~\cite{Entanglement_and_Decoherence,breuer_petruccione_book} in topological systems remains insufficiently explored. 
Crucially, topological systems are not inherently robust to decoherence, raising fundamental questions about the stability of topological order in realistic,  and therefore open, systems.

For the realization of topological quantum computing~\cite{RevModPhys.88.041001,doi:10.1126/science.270.5242.1633,PhysRevLett.85.1762,10.1002/1521-3978,Ladd2010,PhysRevA.51.1015,10.21468/SciPostPhys.3.3.021,RevModPhys.80.1083,Andrew_Steane_1998}, it is crucial to understand the extent to which the topological protection observed in closed systems persists under realistic conditions involving coupling to an environment.
MZMs, predicted in topological superconductors~\cite{Alicea_2012, Sato_2017}, are candidates for realizing topological qubits~\cite{RevModPhys.80.1083, annurev:/content/journals/10.1146/annurev-conmatphys-030212-184337, Alicea2011, 10.21468/SciPostPhysLectNotes.15}.
Topological qubits are generally expected to exhibit a degree of robustness against decoherence. However, while some studies~\cite{PhysRevB.85.165124, PhysRevB.92.165118} support this view, others have demonstrated that MZMs can be susceptible to decoherence, both in general and specifically during braiding operations~\cite{PhysRevB.85.121405, PhysRevB.92.165118, PhysRevB.92.115441, PhysRevLett.115.120402, PhysRevB.84.205109, PhysRevB.85.174533}.
Furthermore, the $B$ phase of the KHM with time-reversal symmetry breaking exhibits localized MZMs at its vortex excitations, indicating its potential for topological quantum computation~\cite{KITAEV20062,RevModPhys.80.1083,KITAEV20032,10.21468/SciPostPhys.3.3.021}.

Quantum spin liquids (QSLs)~\cite{10.1038/nature08917} represent a phase of matter that typically exhibits quantum fluctuations,  frustration, long-range entanglement, fractionalized excitations, emergent gauge fields, and the absence of long-range magnetic order down to zero temperature. The KHM~\cite{KITAEV20062} is a rare example of an exactly solvable, strongly interacting two-dimensional system whose ground state is a topological QSL. This model consists of spin-$1/2$ particles arranged on a honeycomb lattice, interacting via direction-dependent Ising-like couplings. Under open boundary conditions (OBC), the KHM supports edge states that are MZMs~\cite{PhysRevB.89.235434, PhysRevB.99.184418}. Its exact solvability facilitates detailed studies of diverse topological phases and quantum phase transitions, including Abelian, non-Abelian, chiral, and non-chiral phases.

STM has been proposed as a method to probe MZMs in the Kitaev QSL~\cite{PhysRevB.102.134423,PhysRevLett.126.127201,PhysRevLett.125.267206,2411.01784,2501.05608}. However, STM setups inherently involve interactions between the KHM monolayer and the underlying substrate, making it crucial to understand how the substrate affects the properties of the KHM.
We utilize the Lindblad master equation (LME)~\cite{breuer_petruccione_book}  to analyze how the substrate influences the properties of the KHM.

Experimental realization of a Kitaev QSL remains challenging, as the
Kitaev interaction in all materials studied so far is accompanied by
additional competing interactions~\cite{Mandal_2025,
  2501.05608,10.1088/1361-648X/aa8cf5}. Nevertheless, monolayers of
candidate Kitaev QSL materials have been successfully fabricated on
surfaces~\cite{2501.05608,D2NR02827A,2403.16553}. A different route to
MZMs are  atomic spin chains on superconducting substrates,
created using an STM setup~\cite{LoConte2024,RACHEL20251}. A theoretical model
featuring MZMs in one-dimensional topological superconductors is the Kitaev
chain (KC)~\cite{Kitaev_chain_original}. While this appears at
  first sight to be quite a different setup from a QSL, some
  regimes of the KHM can be mapped onto the KC, which we will also
  discuss.

In Sec.~\ref{section:model}, we present the KHM Hamiltonian for a cylindrical geometry and introduce the LME.
In Sec.~\ref{sec:rho_ss}, we discuss the time evolution of the density matrix, with particular focus on the steady state.
Afterward,  we consider the case $J_x=J_y=J_z$ of the KHM (see Eq.~\eqref{eq:model_kitaev_ham_general}). 
First, in  Sec.~\ref{sec:ent_fid}, we analyze the time evolution of the entropy and fidelity.
Second, in Sec.~\ref{sec:E_K_time_evol}, we elaborate on how the band structure evolves.
Third,  Sec.~\ref{sec:tau}, we characterize the time evolution towards the steady state with a relaxation time.
In Sec.~\ref{sec:vary_J_i} we discuss how the previously obtained results change when the coupling constants are not equal anymore. 
Finally, Sec.~\ref{sec:discuss} summarizes our findings and outlines promising directions for future studies. 
 
\section{Model and methods}\label{section:model}

The Hamiltonian for the KHM~\cite{KITAEV20062} is given by
\begin{align}
    H =   \sum_{\alpha \in \{x,y,z\}}   \sum_{\langle i,j \rangle_\alpha}     J_\alpha      \sigma_i^\alpha     \sigma_j^\alpha,
\label{eq:model_kitaev_ham_general}
\end{align}
where the sum $\sum_{\langle i,j \rangle_\alpha}$ runs over all nearest-neighbor pairs connected by bonds of type $\alpha$, and $J_\alpha$ denotes the corresponding coupling strength.
Throughout this work, we adopt the convention $J_x + J_y + J_z = 1$ (with $\hbar = 1$).
This bond-dependent Ising-like interaction can arise in certain materials~\cite{10.1088/1361-648X/aa8cf5,Mandal_2025,2501.05608} via the Jackeli-Khaliullin mechanism~\cite{PhysRevLett.102.017205}, and can lead to effective coupling parameters associated with the $B$ phase of the  KHM.
Accordingly, our primary focus will be on the $B$ phase, and in the following
sections, we will primarily consider $J_x = J_y = J_z$.
However, we will also, in some places, mention phases $A_{i}$ with
$i=x,y,z$ referring to the gapped spin liquids found for  
$J_{i}\gg J_{j\neq i}$. 

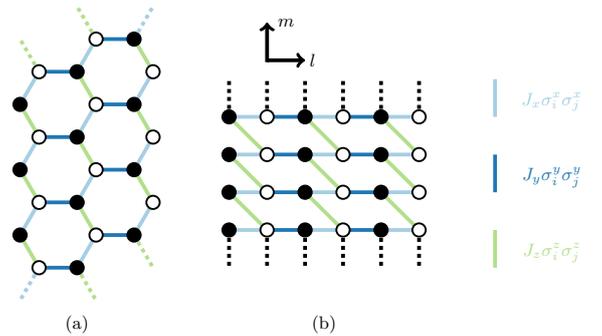
\begin{figure}
\centering
\scalebox{0.5}{
\begin{tikzpicture}[rotate=90]
\definecolor{x_bond_color}{RGB}{166,206,227}
\definecolor{y_bond_color}{RGB}{31,120,180}
\definecolor{z_bond_color}{RGB}{178,223,138}
\definecolor{lat_vec_col}{RGB}{51,160,44}

\draw[line width=1mm,color=y_bond_color] (0,0) to (0,1);
\draw[line width=1mm,color=z_bond_color] (0,1) to (0.866, 1.5);
\draw[line width=1mm,color=x_bond_color] (0.866, 1.5) to (1.732, 1);
\draw[line width=1mm,color=y_bond_color] (1.732,0) to (1.732,1);
\draw[line width=1mm,color=x_bond_color] (0, 0) to (0.866, -0.5);
\draw[line width=1mm,color=z_bond_color] (0.866, -0.5) to (1.732,0);
\draw[line width=1mm,color=z_bond_color] (1.732,1) to (2.598, 1.5);
\draw[line width=1mm,color=x_bond_color] (2.598, 1.5) to (3.464, 1);
\draw[line width=1mm,color=y_bond_color] (3.464,0) to (3.464,1);
\draw[line width=1mm,color=x_bond_color] (1.732, 0) to (2.598, -0.5);
\draw[line width=1mm,color=z_bond_color] (2.598, -0.5) to (3.464,0);
\draw[line width=1mm,color=z_bond_color] (3.464,1) to (4.33013, 1.5);
\draw[line width=1mm,color=x_bond_color] (4.33013, 1.5) to (5.19615, 1);
\draw[line width=1mm,color=y_bond_color] (5.19615,0) to (5.19615,1);
\draw[line width=1mm,color=x_bond_color] (3.464, 0) to (4.33013, -0.5);
\draw[line width=1mm,color=z_bond_color] (4.33013, -0.5) to (5.19615,0);
\draw[line width=1mm,color=y_bond_color]  (0.866, -0.5) to (0.866, -1.5);
\draw[line width=1mm,color=x_bond_color] (0.866, -1.5) to (1.732, -2);
\draw[line width=1mm,color=z_bond_color]  (1.732, -2) to  (2.598, -1.5);
\draw[line width=1mm,color=y_bond_color]   (2.598, -1.5) to (2.598, -0.5);
\draw[line width=1mm,color=x_bond_color]    (2.598, -1.5) to (3.464, -2) ;
\draw[line width=1mm,color=z_bond_color]    (3.464, -2) to (4.33013, -1.5);
\draw[line width=1mm,color=y_bond_color]    (4.33013, -1.5) to  (4.33013, -0.5);
\draw[line width=1mm,color=x_bond_color]    (4.33013, -1.5) to (5.19615,-2);
\draw[line width=1mm,color=z_bond_color]    (5.19615,-2) to (6.06218,-1.5) ;
\draw[line width=1mm,color=y_bond_color]    (6.06218,-1.5) to (6.06218,-0.5);
\draw[line width=1mm,color=x_bond_color]     (6.06218,-0.5) to (5.19615,0) ;

\draw[dashed,line width=1mm,color=z_bond_color] (0,0) to (-0.866,-0.5);
\draw[dashed,line width=1mm,color=x_bond_color] (0,1) to (-0.866,1.5);

\draw[dashed,line width=1mm,color=z_bond_color] (5.196153,1) to (6.06218,1.5);
\draw[dashed,line width=1mm,color=z_bond_color] (6.06218,-0.5) to (6.9282,0);
\draw[dashed,line width=1mm,color=x_bond_color] (6.06218,-1.5) to (6.9282,-2);
\draw[dashed,line width=1mm,color=z_bond_color] (0.866,-1.5) to (0,-2);

\node[circle,draw=black, line width=0.5mm, fill=black, inner sep=0pt,minimum size=10pt] (b) at (0,0) {};
\node[circle,draw=black, line width=0.5mm, fill=black, inner sep=0pt,minimum size=10pt] (b) at (0.866, 1.5) {};
\node[circle,draw=black, line width=0.5mm, fill=white, inner sep=0pt,minimum size=10pt] (b) at (1.732, 1) {};
\node[circle,draw=black, line width=0.5mm, fill=black, inner sep=0pt,minimum size=10pt] (b) at (1.732,0) {};
\node[circle,draw=black, line width=0.5mm, fill=white, inner sep=0pt,minimum size=10pt] (b) at  (0.866, -0.5)  {};
\node[circle,draw=black, line width=0.5mm, fill=white, inner sep=0pt,minimum size=10pt] (b) at (0,1) {};
\node[circle,draw=black, line width=0.5mm, fill=black, inner sep=0pt,minimum size=10pt] (b) at (0.866,-1.5) {};
\node[circle,draw=black, line width=0.5mm, fill=white, inner sep=0pt,minimum size=10pt] (b) at (1.732,-2) {};
\node[circle,draw=black, line width=0.5mm, fill=black, inner sep=0pt,minimum size=10pt] (b) at (2.598,-1.5) {};
\node[circle,draw=black, line width=0.5mm, fill=white, inner sep=0pt,minimum size=10pt] (b) at (2.598,-0.5) {};
\node[circle,draw=black, line width=0.5mm, fill=black, inner sep=0pt,minimum size=10pt] (b) at (2.598,1.5) {};
\node[circle,draw=black, line width=0.5mm, fill=white, inner sep=0pt,minimum size=10pt] (b) at (3.464,1) {};
\node[circle,draw=black, line width=0.5mm, fill=black, inner sep=0pt,minimum size=10pt] (b) at (3.464,0) {};
\node[circle,draw=black, line width=0.5mm, fill=white, inner sep=0pt,minimum size=10pt] (b) at (3.464,-2) {};
\node[circle,draw=black, line width=0.5mm, fill=black, inner sep=0pt,minimum size=10pt] (b) at (4.33013,-1.5) {};
\node[circle,draw=black, line width=0.5mm, fill=white, inner sep=0pt,minimum size=10pt] (b) at (4.33013,-0.5) {};
\node[circle,draw=black, line width=0.5mm, fill=black, inner sep=0pt,minimum size=10pt] (b) at (4.33013,1.5) {};
\node[circle,draw=black, line width=0.5mm, fill=white, inner sep=0pt,minimum size=10pt] (b) at (5.19615,-2) {};
\node[circle,draw=black, line width=0.5mm, fill=black, inner sep=0pt,minimum size=10pt] (b) at (5.19615,-0) {};
\node[circle,draw=black, line width=0.5mm, fill=white, inner sep=0pt,minimum size=10pt] (b) at (5.19615,1) {};
\node[circle,draw=black, line width=0.5mm, fill=black, inner sep=0pt,minimum size=10pt] (b) at (6.06218,-1.5) {};
\node[circle,draw=black, line width=0.5mm, fill=white, inner sep=0pt,minimum size=10pt] (b) at (6.06218,-0.5) {};

\draw[line width=1mm,color=x_bond_color] (1, -4) to (1, -5);
\draw[line width=1mm,color=y_bond_color] (1, -5) to (1, -6);
\draw[line width=1mm,color=x_bond_color] (1, -6) to (1, -7);
\draw[line width=1mm,color=y_bond_color] (1, -7) to (1, -8);
\draw[line width=1mm,color=x_bond_color] (1, -8) to (1, -9);

\draw[line width=1mm,color=x_bond_color] (2, -4) to (2, -5);
\draw[line width=1mm,color=y_bond_color] (2, -5) to (2, -6);
\draw[line width=1mm,color=x_bond_color] (2, -6) to (2, -7);
\draw[line width=1mm,color=y_bond_color] (2, -7) to (2, -8);
\draw[line width=1mm,color=x_bond_color] (2, -8) to (2, -9);

\draw[line width=1mm,color=x_bond_color] (3, -4) to (3, -5);
\draw[line width=1mm,color=y_bond_color] (3, -5) to (3, -6);
\draw[line width=1mm,color=x_bond_color] (3, -6) to (3, -7);
\draw[line width=1mm,color=y_bond_color] (3, -7) to (3, -8);
\draw[line width=1mm,color=x_bond_color] (3, -8) to (3, -9);

\draw[line width=1mm,color=x_bond_color] (4, -4) to (4, -5);
\draw[line width=1mm,color=y_bond_color] (4, -5) to (4, -6);
\draw[line width=1mm,color=x_bond_color] (4, -6) to (4, -7);
\draw[line width=1mm,color=y_bond_color] (4, -7) to (4, -8);
\draw[line width=1mm,color=x_bond_color] (4, -8) to (4, -9);

\draw[line width=1mm,color=z_bond_color] (1,-5) to (2,-4);
\draw[line width=1mm,color=z_bond_color] (1,-7) to (2,-6);
\draw[line width=1mm,color=z_bond_color] (1,-9) to (2,-8);

\draw[line width=1mm,color=z_bond_color] (2,-5) to (3,-4);
\draw[line width=1mm,color=z_bond_color] (2,-7) to (3,-6);
\draw[line width=1mm,color=z_bond_color] (2,-9) to (3,-8);

\draw[line width=1mm,color=z_bond_color] (3,-5) to (4,-4);
\draw[line width=1mm,color=z_bond_color] (3,-7) to (4,-6);
\draw[line width=1mm,color=z_bond_color] (3,-9) to (4,-8);

\draw[dashed, line width=1mm,color=black] (1, -9) to (0, -9);
\draw[dashed, line width=1mm,color=black] (1, -8) to (0, -8);
\draw[dashed, line width=1mm,color=black] (1, -7) to (0, -7);
\draw[dashed, line width=1mm,color=black] (1, -6) to (0, -6);
\draw[dashed, line width=1mm,color=black] (1, -5) to (0, -5);
\draw[dashed, line width=1mm,color=black] (1, -4) to (0, -4);
\draw[dashed, line width=1mm,color=black] (4, -9) to (5, -9);
\draw[dashed, line width=1mm,color=black] (4, -8) to (5, -8);
\draw[dashed, line width=1mm,color=black] (4, -7) to (5, -7);
\draw[dashed, line width=1mm,color=black] (4, -6) to (5, -6);
\draw[dashed, line width=1mm,color=black] (4, -5) to (5, -5);
\draw[dashed, line width=1mm,color=black] (4, -4) to (5, -4);

\node[circle,draw=black, line width=0.5mm, fill=black, inner sep=0pt,minimum size=10pt] (b) at (1, -4) {};
\node[circle,draw=black, line width=0.5mm, fill=white, inner sep=0pt,minimum size=10pt] (b) at (1, -5) {};
\node[circle,draw=black, line width=0.5mm, fill=black, inner sep=0pt,minimum size=10pt] (b) at (1, -6) {};
\node[circle,draw=black, line width=0.5mm, fill=white, inner sep=0pt,minimum size=10pt] (b) at (1, -7) {};
\node[circle,draw=black, line width=0.5mm, fill=black, inner sep=0pt,minimum size=10pt] (b) at (1, -8) {};
\node[circle,draw=black, line width=0.5mm, fill=white, inner sep=0pt,minimum size=10pt] (b) at (1, -9) {};

\node[circle,draw=black, line width=0.5mm, fill=black, inner sep=0pt,minimum size=10pt] (b) at (2, -4) {};
\node[circle,draw=black, line width=0.5mm, fill=white, inner sep=0pt,minimum size=10pt] (b) at (2, -5) {};
\node[circle,draw=black, line width=0.5mm, fill=black, inner sep=0pt,minimum size=10pt] (b) at (2, -6) {};
\node[circle,draw=black, line width=0.5mm, fill=white, inner sep=0pt,minimum size=10pt] (b) at (2, -7) {};
\node[circle,draw=black, line width=0.5mm, fill=black, inner sep=0pt,minimum size=10pt] (b) at (2, -8) {};
\node[circle,draw=black, line width=0.5mm, fill=white, inner sep=0pt,minimum size=10pt] (b) at (2, -9) {};

\node[circle,draw=black, line width=0.5mm, fill=black, inner sep=0pt,minimum size=10pt] (b) at (3, -4) {};
\node[circle,draw=black, line width=0.5mm, fill=white, inner sep=0pt,minimum size=10pt] (b) at (3, -5) {};
\node[circle,draw=black, line width=0.5mm, fill=black, inner sep=0pt,minimum size=10pt] (b) at (3, -6) {};
\node[circle,draw=black, line width=0.5mm, fill=white, inner sep=0pt,minimum size=10pt] (b) at (3, -7) {};
\node[circle,draw=black, line width=0.5mm, fill=black, inner sep=0pt,minimum size=10pt] (b) at (3, -8) {};
\node[circle,draw=black, line width=0.5mm, fill=white, inner sep=0pt,minimum size=10pt] (b) at (3, -9) {};

\node[circle,draw=black, line width=0.5mm, fill=black, inner sep=0pt,minimum size=10pt] (b) at (4, -4) {};
\node[circle,draw=black, line width=0.5mm, fill=white, inner sep=0pt,minimum size=10pt] (b) at (4, -5) {};
\node[circle,draw=black, line width=0.5mm, fill=black, inner sep=0pt,minimum size=10pt] (b) at (4, -6) {};
\node[circle,draw=black, line width=0.5mm, fill=white, inner sep=0pt,minimum size=10pt] (b) at (4, -7) {};
\node[circle,draw=black, line width=0.5mm, fill=black, inner sep=0pt,minimum size=10pt] (b) at (4, -8) {};
\node[circle,draw=black, line width=0.5mm, fill=white, inner sep=0pt,minimum size=10pt] (b) at (4, -9) {};

\draw[->, line width=1mm] (5.5,-5) to (5.5,-6);
\draw[->, line width=1mm] (5.45,-5) to (6.5,-5);
\node[scale=1.5] at(5.5,-6.2) {$l$};
\node[scale=1.5] at(6.5,-5.5) {$m$};

\draw[line width=1mm,color=x_bond_color]     (4,-11) to (5,-11);
\node[color=x_bond_color, scale=1.5] at (4.4, -12.5) {$J_x     \sigma_i^x     \sigma_j^x$};
\draw[line width=1mm,color=y_bond_color]   (2,-11) to (3,-11);
\node[color=y_bond_color,scale=1.5] at (2.4, -12.5) {$J_y     \sigma_i^y     \sigma_j^y$};
\draw[line width=1mm,color=z_bond_color]     (0,-11) to (1,-11);
\node[color=z_bond_color, scale=1.5] at (0.4, -12.5) {$J_z     \sigma_i^z     \sigma_j^z$};

\node[color=black, scale=1.5] at (-1.5,-6.5) {(b)};
\node[color=black, scale=1.5] at (-1.5,0) {(a)};

\end{tikzpicture}}
\caption{This figure depicts two lattice structures, the KHM lattice (a) and the deformed brick-wall lattice (b), which are topologically equivalent. PBC are imposed along the $m$ direction, indicated by dashed lines, while OBC are applied along the orthogonal $l$ direction. The lattice sites are distinguished by their $l$ values: black circles denote sites with odd $l$, whereas white circles denote sites with even $l$.}
\label{fig:model_kitaev_half_open_sketch}
\end{figure}

Using a cylindrical geometry with zigzag edges in the KHM (see
Fig.~\ref{fig:model_kitaev_half_open_sketch}) takes advantage of its
exact solvability to make calculations easier and allow MZMs to
appear. 
The detailed derivation of the KHM Hamiltonian for this geometry and
the flux-free sector can be found in Ref.~\cite{PhysRevB.99.184418}. 
The resulting Hamiltonian is
\begin{equation}
\begin{aligned}
  H&= \sum_{l=1}^{L}  \sum_{k} \Big( J_x-J_z \cos(k) \Big)   \left[  \alpha_{(l,k)}^\dagger   \alpha_{(l,k)}^{\vphantom{\dagger}}  -   \alpha_{(l,-k)}^{\vphantom{\dagger}}    \alpha_{(l,-k)}^\dagger   \right] \\
&+ \frac{J_y}{2} \sum_{l=1}^{L-1}  \sum_{k} \left[    \alpha_{(l,-k)}^{\vphantom{\dagger}}  \alpha_{(l+1,k)}^{\vphantom{\dagger}} -   \alpha_{(l+1,-k)}^{\vphantom{\dagger}}   \alpha_{(l,k)}^{\vphantom{\dagger}}  \color{white} \right] \\
& \hspace{0.4cm}  \normalcolor -   \alpha_{(l,k)}^\dagger   \alpha_{(l+1,-k)}^\dagger  +    \alpha_{(l+1,k)}^\dagger    \alpha_{(l,-k)}^\dagger  +    \alpha_{(l,-k)}^{\vphantom{\dagger}}  \alpha_{(l+1,-k)}^\dagger  \\
 & \hspace{0.25cm} \color{white} \left[ \normalcolor  -  \alpha_{(l+1,k)}^\dagger   \alpha_{(l,k)}^{\vphantom{\dagger}} -   \alpha_{(l,k)}^\dagger   \alpha_{(l+1,k)}^{\vphantom{\dagger}} +   \alpha_{(l+1,-k)}^{\vphantom{\dagger}}   \alpha_{(l,-k)}^\dagger \right] \\
& + J_z \sum_{l=1}^{L} \sum_{k} i  \sin(k) \left[   \alpha_{(l,-k)}^{\vphantom{\dagger}}   \alpha_{(l,k)}^{\vphantom{\dagger}} -    \alpha_{(l,k)}^\dagger   \alpha_{(l,-k)}^\dagger \right],
\label{eq:result_KMH_HOBC_ham}
\end{aligned}
\end{equation}
where $L$ is the number of sites along the OBC direction, $l$ indexes
these sites,  and  $\alpha_{(l,k)}$  is  the complex fermionic
operator at site $l$.

This fermionic formulation of the KHM also illustrates that one can
use the same model to investigate the KC. Setting $J_z = 0$ gives us
KCs with OBC, while restricting the length of the cylinder to $L = 1$
leads to the KC with periodic boundary conditions (PBC). The same scenario can also be achieved by
setting $J_y=0$, which yields $L$ identical copies of the KC with PBC.

We use the Bogoliubov-de Gennes (BdG) formalism to obtain the
bands of the closed KHM. Due to the redundancy arising from the
introduction of hole operators, we exclude bands with negative
energy to avoid double counting. Remaining bands are then labeled in
ascending order, so that  $n_\mathrm{band} = 1$ corresponds to
the lowest band with positive energy. 

During the derivation of this Hamiltonian in Ref.~\cite{PhysRevB.99.184418}, the fact that the ground state lies in the flux-free sector was exploited to eliminate quartic terms.
This simplification raises questions about its suitability for calculating time evolution via the LME.
While time evolution based on the LME generally requires considering the entire Hilbert space, extending beyond the flux-free sector introduces quartic terms.
Based on Refs.~\cite{PhysRevLett.126.077201, PhysRevResearch.3.L032024, 2208.07732}, restricting the analysis to the flux-free sector may be a reasonable approximation, and we, therefore, adopt this simplest approach.

The presence of MZMs depends on both the coupling strengths and the
edge structure, specifically on whether the geometry is zigzag or
armchair~\cite{PhysRevB.89.235434, PhysRevB.99.184418}.  
For the zigzag edge geometry analyzed here, the  KHM exhibits MZMs in both the B phase and the $A_y$ phase. 
These MZMs form a flat band at zero energy~\cite{PhysRevB.103.184507},
however, modes at different edges interact for finite systems, leading
to the finite-size effect of a nonzero energy, which decreases
exponentially with increasing system size $L$. 

To accurately model an STM setup, the interaction between the KHM monolayer and the surface must be taken into account. 
We employ an effective model that describes the KHM as an open quantum system interacting with the surface, which acts as its environment.
The dynamics of open quantum systems are commonly described by master equations~\cite{breuer_petruccione_book}.
Lindblad~\cite{lindblad_derivation} and Gorini, Kossakowski, and Sudarshan~\cite{doi:10.1063/1.522979} derived the general structure of the generator $\mathcal{L}$, which describes the Markovian time evolution of density matrices.
The LME is given by
\begin{align}
\dot{\rho}=\mathcal{L}(\rho) = i [H,\rho] +  \gamma  \sum_i  \left( L_i^{\vphantom{\dagger}} \rho L_i^{\dagger} -\frac{1}{2} \Big\{\ L_i^{\dagger} L_i^{\vphantom{\dagger}}, \rho \Big\}\   \right),
\label{eq:lindblad_LME}
\end{align}
where $L_i$ are the jump operators, $\gamma>0$  the dissipation strength, and $\hbar=1$.  
As an initial condition, we assume that the coupling between the system and its environment is switched on at $t = 0$.  Derivations of the LME can be found in Refs.~\cite{Entanglement_and_Decoherence,PhysRevB.104.094306,breuer_petruccione_book,lindblad_derivation,doi:10.1063/1.522979,brasil.fanchini.napolitano.2013,1902.00967,10.1063/1.5115323}. Details of the numerical methods employed are provided in Sec.~\ref{appendix:LME}.
The LME is employed to determine the time evolution of the eigenstates of the Hamiltonian in Eq.~\eqref{eq:result_KMH_HOBC_ham}. 

The aim is to compute the time evolution using the LME, with a focus
on decoherence. The Pauli matrix $ \sigma^z$  is known to induce
decoherence~\cite{breuer_petruccione_book,PhysRevE.92.042143,PhysRevLett.111.150403,PhysRevA.97.052106,PhysRevB.99.174303,PhysRevE.83.011108}. After
transformation with the same Jordan-Wigner
transformation (JWT) applied during the derivation of the
Hamiltonian~\cite{PhysRevB.99.184418}, it corresponds to the number
operator (up to factors), which then induces decoherence in the
fermionic
system~\cite{Li_2012,PhysRevLett.117.137202,PhysRevLett.122.040604,PhysRevB.102.100301,PRXQuantum.2.030350}.
The complete list of jump operators examined is in Tab.~\ref{tab:KHM_list_jump_operators}.
We employ one jump operator for each of the  $L$
sites along the OBC direction, with a uniform dissipation rate
$\gamma$ for each.

\section{Results} \label{sec:results}

We first discuss the time evolution of density matrices in
Sec.~\ref{sec:rho_ss}, then go to quantities like entropy and
fidelity in Sec.~\ref{sec:ent_fid}, and finally discuss band structure and the spectral
gap in Secs.~\ref{sec:E_K_time_evol} and~\ref{sec:tau}. We finally
vary parameter of the Hamiltonian in Sec.~\ref{sec:vary_J_i}.

\subsection{Time evolution of density matrices -- steady states}\label{sec:rho_ss}
In this section, we discuss the time evolution of density matrices.  
The detailed features for the different jump operators,  such as whether
they cause decoherence, preserve the BdG-type particle-hole symmetry
(PHS), and their steady states are discussed in
Sec.~\ref{sec_appendix_jump_operators}.

The steady state depends on $k$, $J_\mathrm{\alpha}$, $L$, and the jump
operators (see Tab.~\ref{tab:KHM_list_steady_state}). Remarkably,
fundamentally different sets of jump operators can yield the same steady
state. Often, this steady state is the maximally mixed (mm) state, 
\begin{align}  
\rho_\mathrm{mm} = \frac{1}{2L} \mathbb{1},  
\label{eq:kitaev_max_mixed_state}  
\end{align}  
where $\mathbb{1}$ is the identity matrix. 
If $\rho_\mathrm{mm}$ is the steady state, it is unique, and this steady
states can arise from quite different jump operators, for instance involving number
operators, pair creation and annihilation operators, and incoherent hopping
operators (see Tab.~\ref{tab:KHM_list_steady_state}). 
Although these jump operators, and also different parameters for the
Hamiltonian, can lead to distinct short-time dynamics, the long-time
behavior is largely similar due to the common steady state. 

The steady states are identical for all sets of jump operators based on number
operators, although the density matrices differ over time (see
Tab.~\ref{tab:KHM_list_steady_state}). 
This leads to several conclusions.
First, the steady state is independent of whether the BdG-type PHS is
preserved ($L_n$ and $L_{n_{-k}}^{n_{+k}}$) or broken ($L_{n_{+k}}$ and
$L_{n_{-k}}$). 
Second, for the steady state, it does not matter if the jump operators act on the entire Nambu spinor ($L_n$ and $L_{n_{-k}}^{n_{+k}}$) or only half of the Nambu spinor ($L_{n_{+k}}$ and $L_{n_{-k}}$).
Third, it does not matter whether we treat number operators with $\pm k$
via two jump operators per site ($L_{n_{-k}}^{n_{+k}}$) or use a single jump
operator per site ($L_n$).

In addition to sharing a steady state, those jump operators
  that preserve the BdG-type PHS also have quite similar short-time evolution.  
Consequently, as long as $\rho_\mathrm{mm}$ is the steady state (see Tab.~\ref{tab:KHM_list_steady_state}), the time evolution is similar for $L_n$, $L_{n_{-k}}^{n_{+k}}$, $L_{\mathrm{p}}$, $L_{\mathrm{pc}}^{\mathrm{pa}}$, and incoherent jump operators. 
Remarkably, this implies that the time evolution is comparable for
microscopically quite different jump operators based on number operators, pair creation and annihilation operators,
as well as incoherent hopping operators.

While the differences between these sets of jump operators are minimal for the
most symmetric case $J_x = J_y = J_z$, they can become larger for the three $A_i$
phases. A general trend is that as the
disparity between the coupling constants $J_\mathrm{\alpha}$ increases, so do
the differences between the dynamics induced by various sets of jump
operators. Another trend is that differences also increase with 
increasing $\gamma$, however, they typically
remain small as long as $\gamma < J_\alpha$.

\begin{figure}
  \includegraphics[width=\columnwidth]{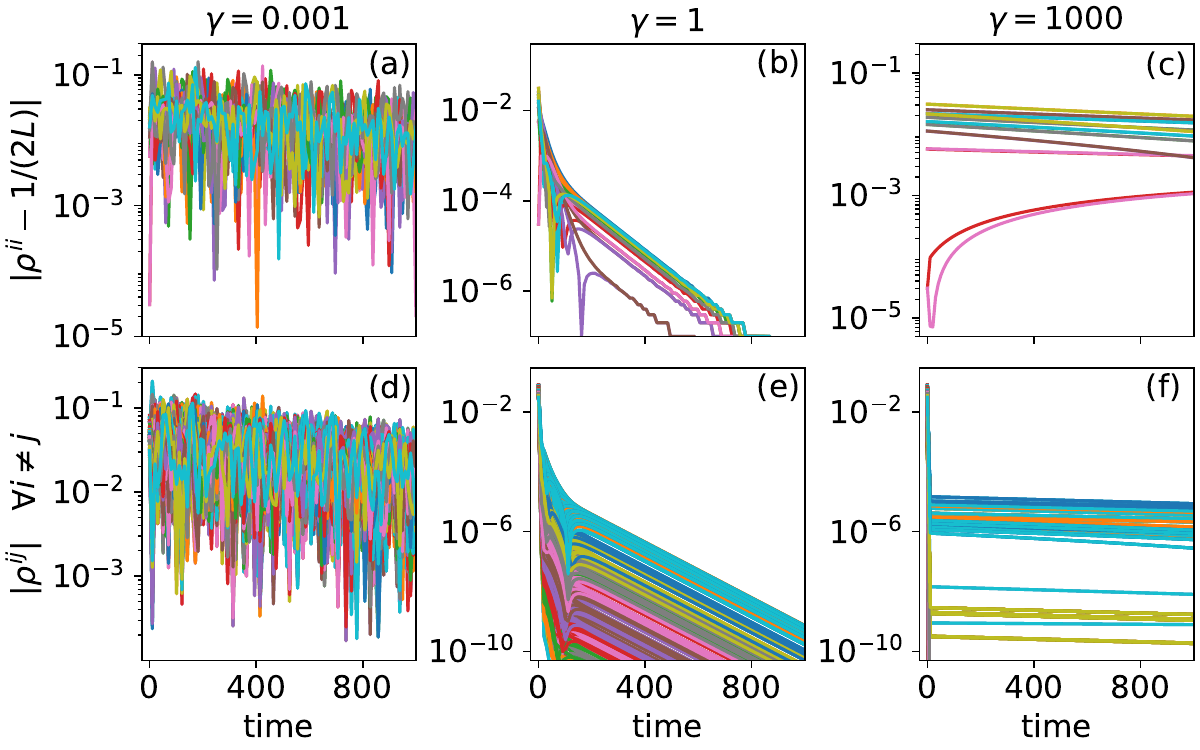}
    \caption{Time evolution of the off-diagonal (bottom row) and diagonal (top row) elements of the density matrix $\rho$, shown relative to the steady state $\rho_{\mathrm{mm}}$. The system parameters are $L = 10$, $J_x = J_y = J_z = 1/3$, $k = 1$, $n_\mathrm{band} = 2$, and the jump operator $L_{n_{-k}}^{n_{+k}}$. The left, middle, and right columns correspond to $\gamma = 0.001$, $\gamma = 1$, and $\gamma = 1000$, respectively. The QZE is observable.
 \label{fig:open_kitaev_rho_QZE_diag}}
\end{figure}

On  the other hand, steady states can differ between OBC and PBC for
the parameter regimes modeling the KC. This distinction is
evident in Tab.~\ref{tab:KHM_list_steady_state} by comparing the steady states
for $J_z = 0$ (KC with OBC), for $L = 1$ (KC with PBC), and for $J_y=0$ ($L$ identical copies of the KC with
PBC).

We find that decoherence can occur independently of the preservation of BdG-type
PHS, as detailed in Tab.~\ref{tab:KHM_list_jump_operators}. Specifically, all
combinations are possible: decoherence with preservation of BdG-type PHS,
decoherence with broken BdG-type PHS, absence of decoherence with preservation
of BdG-type PHS, and absence of decoherence with broken BdG-type PHS. 

Time evolution of density-matrix elements shows a  quantum Zeno effect
(QZE)~\cite{breuer_petruccione_book,10.1063/1.523304} for  every set of
jump operators considered. 
This phenomenon is illustrated in
Fig.~\ref{fig:open_kitaev_rho_QZE_diag}, where it can be observed by
comparing the different columns. Comparison between the left and
middle columns demonstrates that increasing $\gamma$ accelerates the
approach to the steady state. Conversely, comparison between the
middle and right columns shows that further increasing $\gamma$ slows
convergence again. The fastest relaxation to the steady state for $J_x =
J_y = J_z$ occurs roughly when $J_\mathrm{\alpha} \sim \gamma$. As the
differences among the couplings $J_\mathrm{\alpha}$ increase, the
position of this minimum shifts. 

\begin{figure}
  \includegraphics[width=\columnwidth]{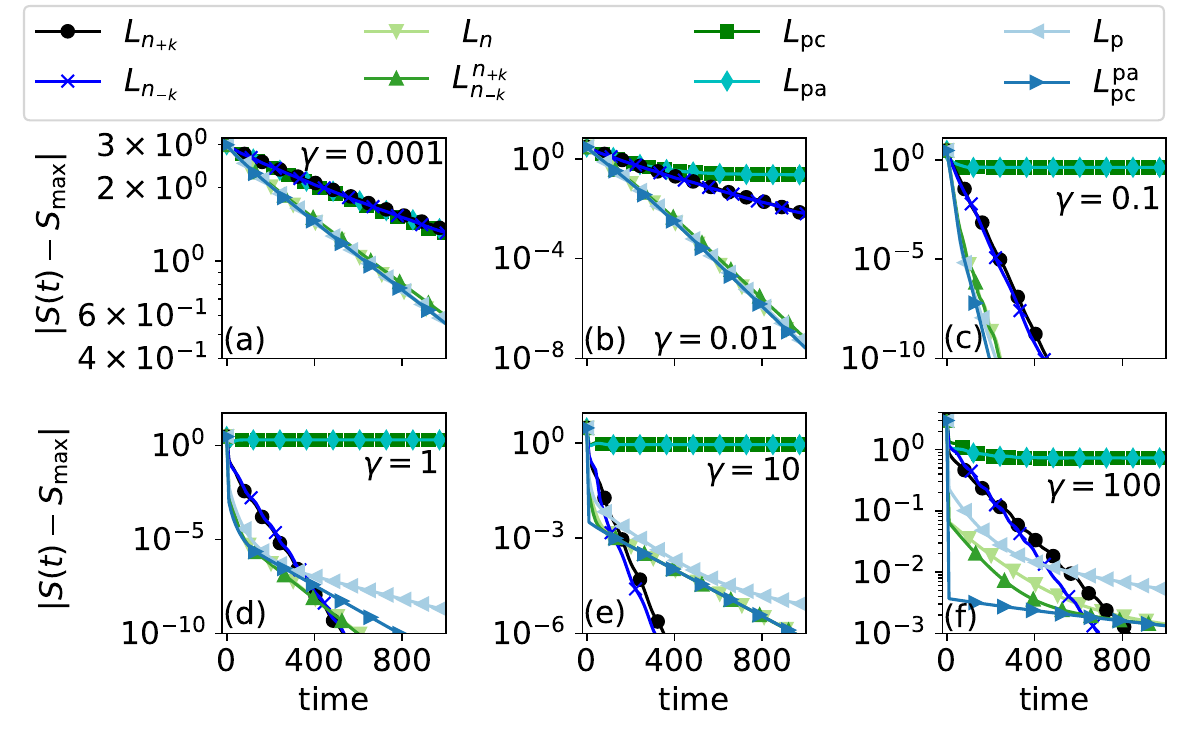}
    \caption{ The time evolution of the entropy for different jump operators  (definitions in Tab.~\ref{tab:KHM_list_jump_operators}), where each panel corresponds to a different $\gamma$. The entropy is shifted by $S_\mathrm{max}$, the steady-state entropy when the system reaches the state $\rho_\mathrm{mm}$ (see Eq.~\eqref{eq:KHM_entropy_SS_mm}). The parameters are $L = 10$, $J_x = J_y = J_z = 1/3$, $n_\mathrm{band} = 2$, and $k = 1$.  
The steady states are listed in Tab.~\ref{tab:KHM_list_steady_state}. Unlike the other jump operators shown in this figure, the steady states for $L_\mathrm{pc}$ and $L_\mathrm{pa}$ are not $\rho_\mathrm{mm}$, which explains why $|S(t) - S_\mathrm{max}|$ does not vanish over time. }
 \label{fig:open_entropy_jump}
\end{figure}

\subsection{Entropy and fidelity}\label{sec:ent_fid}
The time evolution of the density matrix, which transitions from an
initial  eigenstate to a typically mixed steady state, can be
characterized by computing the von Neumann entropy 
\begin{align}  \label{eq:KHM_def_S_P}  
S(t) &= - \mathrm{Tr} \big[ \rho(t) \ln \left( \rho(t) \right) \big].
\end{align}  
In this section, we focus exclusively on the most common steady state,
$\rho_{\mathrm{mm}}$ (see Tab.~\ref{tab:KHM_list_steady_state}), which
has maximum achievable entropy 
\begin{align}  
S(\rho_{\mathrm{mm}}) &=  \ln(2L) = S_\mathrm{max}.
\label{eq:KHM_entropy_SS_mm} 
\end{align}  
Consequently, as long as the steady state is $\rho_{\mathrm{mm}}$, the
entropy evolution converges to the same value regardless of the
initial density matrix or jump operator. 
This section discusses how fast this limit is approached with various
coupling to the environment in 
the case $J_x = J_y = J_z$, while Sec.~\ref{sec:vary_J_i} explores the
impact of non-identical coupling parameters. 

\begin{figure}
  \includegraphics[width=\columnwidth]{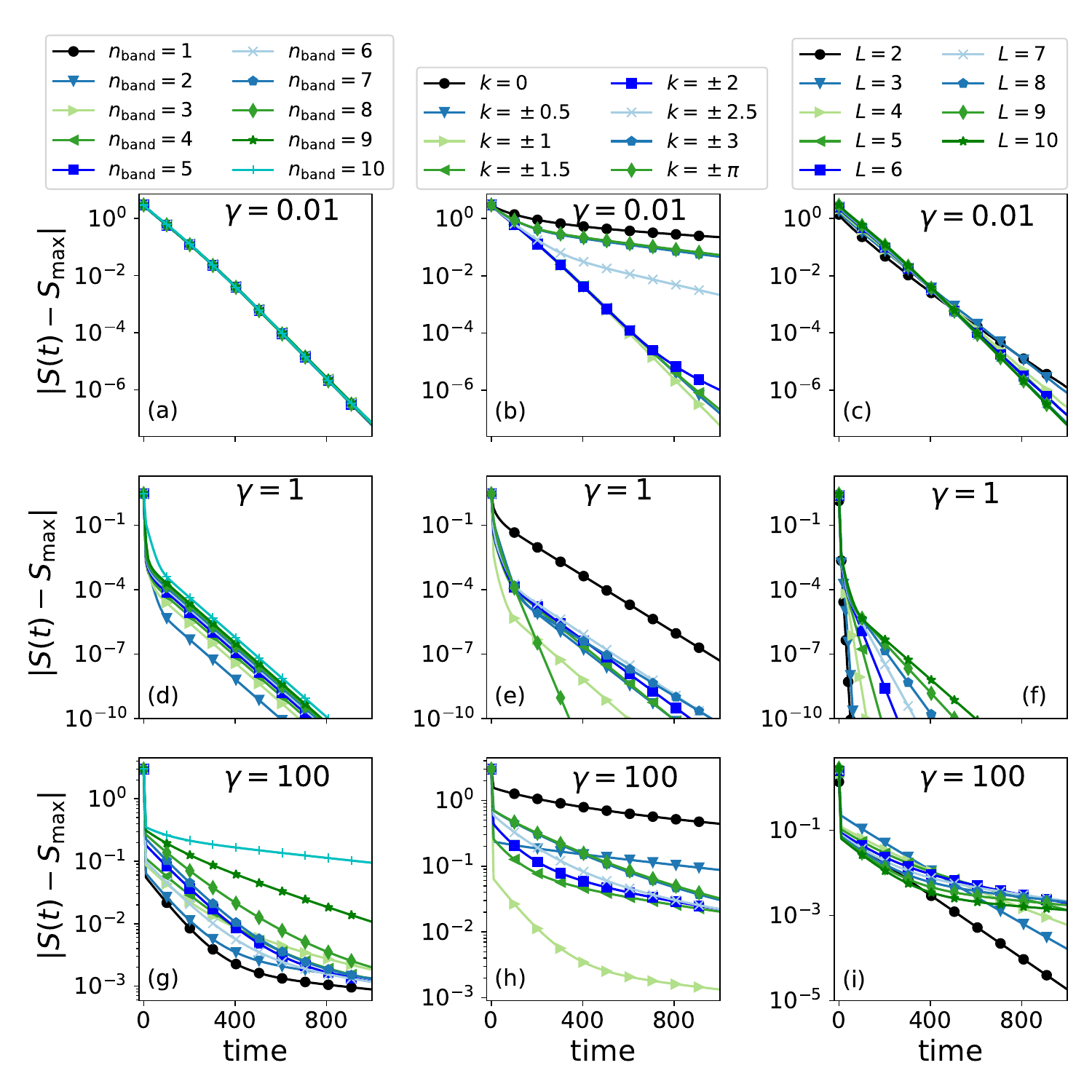}
    \caption{ Time evolution of the entropy, shifted by its steady-state value (see Eq.~\eqref{eq:KHM_entropy_SS_mm}), is shown for different bands (left column), $k$ values (middle column), and finite-size effects (right column). Each row displays the time evolution for a different $\gamma$ value. The parameters, if not varied, are $L = 10$, $J_x = J_y = J_z=1/3$,   $n_\mathrm{band}=2$, $k=1$, and $L_{n_{-k}}^{n_{+k}}$. }
 \label{fig:open_kitaev_entropy_L_10_B_phase_n_k}
\end{figure}

In the case of $\gamma < J_\alpha$, the time evolution of the entropy
is nearly identical for all jump operators from
Tab.~\ref{tab:KHM_list_jump_operators} that preserve the BdG-type PHS,
as shown in Fig.~\ref{fig:open_entropy_jump}.  
This similarity is remarkable, given that the physical processes
described by these jump operators are microscopically quite distinct, and it
is a direct consequence of the similarities in $\rho(t)$ discussed in
Sec.~\ref{sec:rho_ss}.  
While differences emerge for $\gamma > J_\alpha$, they remain weak
compared to the variations in entropy when $k$ changes (see
Fig.~\ref{fig:open_kitaev_entropy_L_10_B_phase_n_k}).

A QZE consistently occurs in $S(t)$, as seen in
Figs.~\ref{fig:open_entropy_jump}
and~\ref{fig:open_kitaev_entropy_L_10_B_phase_n_k}.  
As seen Sec.~\ref{sec:rho_ss} for the $\gamma$-dependence of
$\rho(t)$, the  fastest approach to the steady-state value of the entropy occurs
approximately when $J_\mathrm{\alpha} \sim \gamma$. 
Starting from $J_\mathrm{\alpha} \sim \gamma$, either increasing or
decreasing $\gamma$ causes the time evolution to slow down.    
Furthermore, the kink in $S(t)$ that occurs for $\gamma \gg
J_\mathrm{\alpha}$ is a direct consequence of the kink in $\rho(t)$,
as shown in Fig.~\ref{fig:open_kitaev_rho_QZE_diag}(f). 

While $n_\mathrm{band}$ and $L$ have a negligible impact on the
entropy for $\gamma < J_\alpha$ (see
Figs.~\ref{fig:open_kitaev_entropy_L_10_B_phase_n_k}(a)
and~\ref{fig:open_kitaev_entropy_L_10_B_phase_n_k}(c)), differences
begin to emerge as $\gamma$ increases (see
Figs.~\ref{fig:open_kitaev_entropy_L_10_B_phase_n_k}(g)
and~\ref{fig:open_kitaev_entropy_L_10_B_phase_n_k}(i)).  

In contrast, the entropy shows a strong dependence on $k$ for all
values of $\gamma$ (see
Figs.~\ref{fig:open_kitaev_entropy_L_10_B_phase_n_k}(b),
~\ref{fig:open_kitaev_entropy_L_10_B_phase_n_k}(e)
and~\ref{fig:open_kitaev_entropy_L_10_B_phase_n_k}(h)). 
No general pattern emerges that would allow us to predict which $k$
values lead to a faster or slower increase in entropy relative to
others.

\begin{figure}
  \includegraphics[width=\columnwidth]{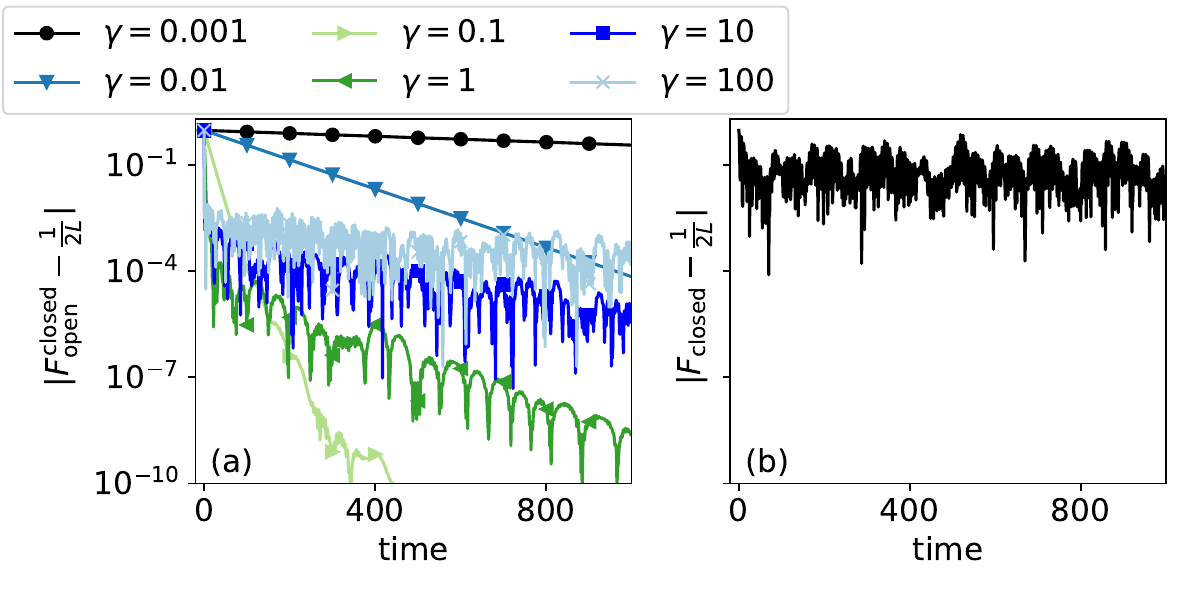}
    \caption{(a) Time evolution of the fidelity defined in Eq.~\eqref{eq:KHM_Fidelity_definitions_open_closed}, shifted by  its steady-state value from Eq.~\eqref{eq:fidelity_SS}, for different values of $\gamma$.
For comparison, (b) shows the fidelity for the unitary time evolution (i.e., $\gamma = 0$) as defined in Eq.~\eqref{eq:KHM_Fidelity_definitions_0_closed}.
The parameters are $J_x = J_y = J_z=1/3 $,  $L_{n_{-k}}^{n_{+k}}$, $L = 10$, $n_\mathrm{band} = 2$, and $k = 1$. 
 \label{fig:open_kitaev_fidelity_def_B_phase}}
\end{figure}

The Uhlmann fidelity $F(t)$ quantifies the similarity between two density matrices:
\begin{subequations}  
\label{eq:KHM_Fidelity_definitions}  
\begin{align}  
F_\mathrm{closed}(t) &= \mathrm{Tr} [\rho(t=0) \cdot \rho_\mathrm{closed}(t)], \label{eq:KHM_Fidelity_definitions_0_closed} \\  
F_\mathrm{open}^\mathrm{closed}(t) &= \mathrm{Tr} [\rho_\mathrm{open}(t) \cdot \rho_\mathrm{closed}(t)], \label{eq:KHM_Fidelity_definitions_open_closed}  
\end{align}  
\end{subequations}  
where $\rho(t=0)$ is the initial density matrix of the closed system, $\rho_\mathrm{open}(t)$ represents the time evolution in an open system ($\gamma > 0$), and $\rho_\mathrm{closed}(t)$ describes the time evolution in a closed system ($\gamma = 0$).
For the most common steady state, $\rho_{\mathrm{mm}}$ (see Tab.~\ref{tab:KHM_list_steady_state}), we obtain
\begin{align}  
F_\mathrm{open}^\mathrm{closed}(\rho_\mathrm{open}=\rho_{\mathrm{mm}}) &=  \frac{1}{2L}. 
\label{eq:fidelity_SS}  
\end{align}  

The KHM with full PBC in both directions was analytically studied in
Ref.~\cite{2208.07732} and an expression for
$F_\mathrm{open}^\mathrm{closed}$ was derived. Additionally,
Ref.~\cite{PhysRevLett.122.040604} calculates the fidelity for various
spin chains and a two-dimensional Bose gas.  
Both references obtain the same analytical expression for the fidelity
under the assumptions of weak coupling ($\gamma \ll J_\alpha$).
Motivated by the analytical form of the fidelity, we investigate
whether our numerically computed fidelities can be accurately
described by such an analytical expression. 
Thus, we use their expression for the fidelity as a fitting function,
adapted by including a finite-size effect term (which depends on the
steady state (ss)): 
\begin{align}
F_\mathrm{fit}(t) = C t^{-b} e^{-at} + \frac{1}{2L} \hspace{1cm} \mathrm{if} \hspace{1cm} \rho_\mathrm{ss} = \rho_\mathrm{mm}.
\label{eq:fit_fidelity}
\end{align} 
We also apply this fitting function to the entropy.

\begin{figure}
  \includegraphics[width=\columnwidth]{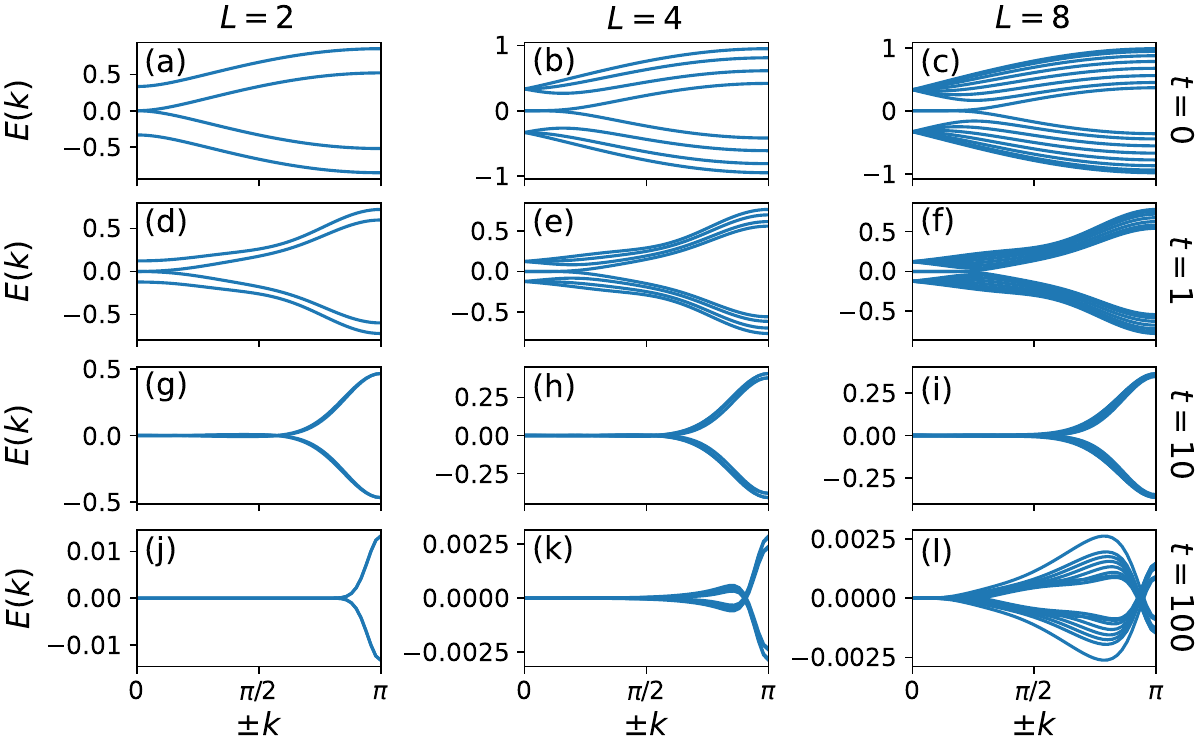}
    \caption{Finite-size effects in the time evolution of the band structure for the parameters $J_x = J_y = J_z = 1/3$, $L_{n_{-k}}^{n_{+k}}$, and $\gamma = 1$. Each column displays the band structure for a specific value of $L$ (above the top row), while each row represents the band structure at a particular time point (to the right of the rightmost column).
 \label{fig:open_kitaev_E_K_B_L}}
\end{figure}

It turns out that this fitting function can be applied to both
$F_\mathrm{open}^\mathrm{closed}$ and the entropy provided that
$\gamma \ll J_\alpha$. Aside from this condition, it is valid for all
parameter regimes, including, in particular, for the MZM. 
This extends previous results, as Ref.~\cite{2208.07732} focuses exclusively on fidelity for the KHM
with PBC in both directions, i.e., without MZMs. 
The necessity of the condition $\gamma \ll J_\alpha$ is  consistent
with the approximations employed in the analytical derivations.  
However, there is no sharp threshold in the ratio $\gamma / J_\alpha$
that delineates where the fitting function becomes invalid. Instead,
fitting errors increase gradually as $\gamma / J_\alpha$ grows, until
a critical value is reached beyond which the fitting procedure fails
to converge. 
Fit parameters vary strongly with the Hamiltonian and LME parameters,
preventing clear or consistent patterns from emerging. 
 
A typical example of the time evolution of the fidelity is shown in
Fig.~\ref{fig:open_kitaev_fidelity_def_B_phase}(a). 
For weak coupling $\gamma \ll J_\alpha$, the fidelity follows a
straight line and becomes, as expected, constant in the limit $\gamma
\to 0$. 
For larger values of $\gamma$, the fidelity no longer follows a
straight line due to variations occurring in the density matrices of
the closed KHM, as shown in
Fig.~\ref{fig:open_kitaev_fidelity_def_B_phase}(b). 
Furthermore, the kink that appears in the density matrices for $\gamma
\gg J_\alpha$ (see Fig.~\ref{fig:open_kitaev_rho_QZE_diag}(f)) also
emerges in the fidelity. 

While the straight lines in
Fig.~\ref{fig:open_kitaev_fidelity_def_B_phase}(a) can be well
described by Eq.~\eqref{eq:fit_fidelity}, once the kink and strong
variations appear, the time evolution becomes more complex than can be
captured by the
analytically derived function of Refs.~\cite{2208.07732,
  PhysRevLett.122.040604}.

\subsection{Time evolution of the band structure}\label{sec:E_K_time_evol}

In this section, we analyze the time evolution of the band structure
for $J_x = J_y = J_z$. A complete list of the jump operators
considered here is provided in Tab.~\ref{tab:KHM_list_jump_operators}.

\begin{figure}
  \includegraphics[width=\columnwidth]{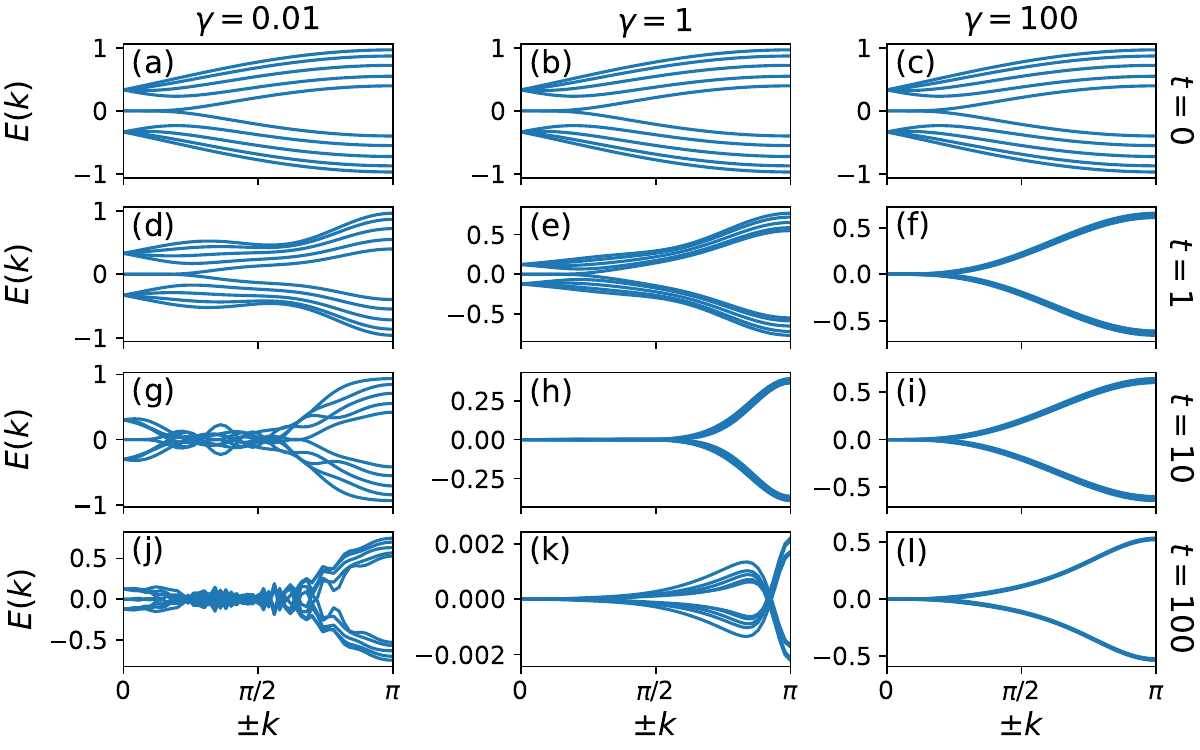}
    \caption{Influence of $\gamma$ on the time evolution of the band structure for $J_x = J_y = J_z = 1/3$, $L = 5$, and $L_{n_{-k}}^{n_{+k}}$. Each column shows the band structure for a specific value of $\gamma$ (above the top row), while each row corresponds to a particular time point (to the right of the rightmost column). The QZE is observable.
 \label{fig:open_kitaev_E_K_B_gamma}}
\end{figure}

As discussed above and summarized in
Tab.~\ref{tab:KHM_list_steady_state}, the steady state is in most
cases $\rho_\mathrm{mm} $. 
The energy associated with $\rho_\mathrm{mm} $ is independent of all
Hamiltonian parameters and is identical across all bands: 
\begin{align}
E(\rho_\mathrm{mm}) = \mathrm{Tr}[\rho_\mathrm{mm} H_\mathrm{KHM}] = 0,
\end{align}
as a consequence of BdG-type PHS. 
In such cases, all energies decay to zero over time. 
Thus, the key remaining questions concern the short-time dynamics and 
the decay rate of the energies. Only in those cases where the steady
state deviates from $\rho_\mathrm{mm}$ do  energies remain nonzero at
all times.

When $\rho_\mathrm{mm}$ is the steady state, the rate at which
energies decay to zero depends sensitively on whether the jump
operators preserve or break the BdG-type PHS. In cases where PHS is
broken (e.g., $L_{n_{+k}}$ or $L_{n_{-k}}$), decay is
significantly slower. This is expected, as such operators act only on
the $+k$ or $-k$ modes, respectively, affecting only half of the BdG
degrees of freedom. In contrast, jump operators that preserve BdG-type
PHS include both $+k$ and $-k$ modes (e.g., $L_{n_{-k}} ^{n_{+k}} $ or
$L_{n}$), acting on all BdG degrees of freedom and facilitating faster
relaxation.

Comparing $L_{n_{+k}}$ and $L_{n_{-k}}$ reveals that their band
structures are related by a sign change in energy. Similarly, the band
structures of $L_\mathrm{pc}$ and $L_\mathrm{pa}$ are connected
through the same energy sign change.

The way operators $L_{n_{+k}}$ and $L_{n_{-k}}$, or $L_{\mathrm{pc}}$ and $L_{\mathrm{pa}}$, which individually break BdG-type PHS, are combined to form PHS preserving jump operators influences the system’s time evolution.
The differences between $L_n$ and $L_{n_{+k}}^{n_{-k}}$ are minimal,
whereas the differences between $L_{\mathrm{pc}}^{\mathrm{pa}}$ and
$L_{\mathrm{p}}$ are more substantial.  
For example, at $t=1$, an energy gap is present for $L_{\mathrm{p}}$,
while no such gap exists for $L_{\mathrm{pc}}^{\mathrm{pa}}$.  
Furthermore, comparing $L_n$ and $L_{n_{+k}}^{n_{-k}}$ with
$L_{\mathrm{pc}}^{\mathrm{pa}}$ and $L_{\mathrm{p}}$ shows that the
energy levels approach $E = 0$ more quickly for
$L_{\mathrm{pc}}^{\mathrm{pa}}$ and $L_{\mathrm{p}}$.

Figure~\ref{fig:open_kitaev_E_K_B_L} shows that finite-size effects
have a weak impact on the time evolution. The bandwidth remains nearly
unchanged between $L = 4$ and $L = 8$, indicating that the approach of
energy levels toward zero is only weakly dependent on system size.  
The bottom row of the figure highlights the emergence of an energy gap
at specific $k$ values. As $L$ increases, this gap becomes apparent at
a growing number of $k$ points.

The ratio $J_\alpha/\gamma$ significantly impacts the time evolution
of the band structure, as shown in
Fig.~\ref{fig:open_kitaev_E_K_B_gamma}. The QZE consistently occurs,
as evidenced by the faster reduction of the energies for $\gamma = 1$
compared to $\gamma = 100$. 
The fastest approach to $E = 0$ occurs, similar to observations made
above,  when $\gamma \sim J_\alpha$. For $\gamma \ll J_\alpha$, time evolution closely
resembles that of the closed system, with one crucial difference: the
energies gradually decrease toward zero. As $\gamma$ increases toward
$J_\alpha$, the energies approach $E = 0$ more rapidly. 
In the regime $\gamma \gg J_\alpha$, the time evolution again
approximates that of the closed system, but with two key differences:  
First, the energies decrease very slowly toward the steady-state energy. 
Second, the energies of the different bands become nearly identical
immediately after coupling to the environment is switched on. This
remains essentially the only change over long timescales, as seen by
the similarity between the band structures at $t = 1$ and $t = 100$. 
In contrast, for $\gamma \ll J_\alpha$, the energy bands remain
well-separated over longer times. The bandwidth is nearly identical in
both the $\gamma \ll J_\alpha$ and $\gamma \gg J_\alpha$ regimes. 

\subsection{Spectral gap}\label{sec:tau}

The long-term behavior can be characterized by relaxation times.
For the LME, the eigenvalues $\lambda_\alpha$ of the $\mathcal{L}$ can be used to determine a relaxation time, since these eigenvalues appear in the exponential function of Eq.~\eqref{eq:LME_vec_solution}.
The real parts $\mathrm{Re}(\lambda_\alpha)$ indicate the rate at
which the steady state is approached. 
The long-time dynamics are governed by the slowest exponential decay,
which directly leads to 
\begin{align}
\frac{1}{\tau}= \Delta\coloneqq - \max\limits_{\substack{\alpha
    \\ \text{Re}[\lambda_\alpha]\neq 0}}{\mathrm{Re}(\lambda_\alpha)},  
\label{eq:def_spectral_gap}
\end{align}
where the inverse of the spectral gap $\Delta$ defines the relaxation
time
$\tau$~\cite{PhysRevB.99.174303,PhysRevLett.111.150403,PhysRevA.89.022118,PhysRevE.92.042143,PhysRevLett.132.070402}.

\begin{figure}
  \includegraphics[width=\columnwidth]{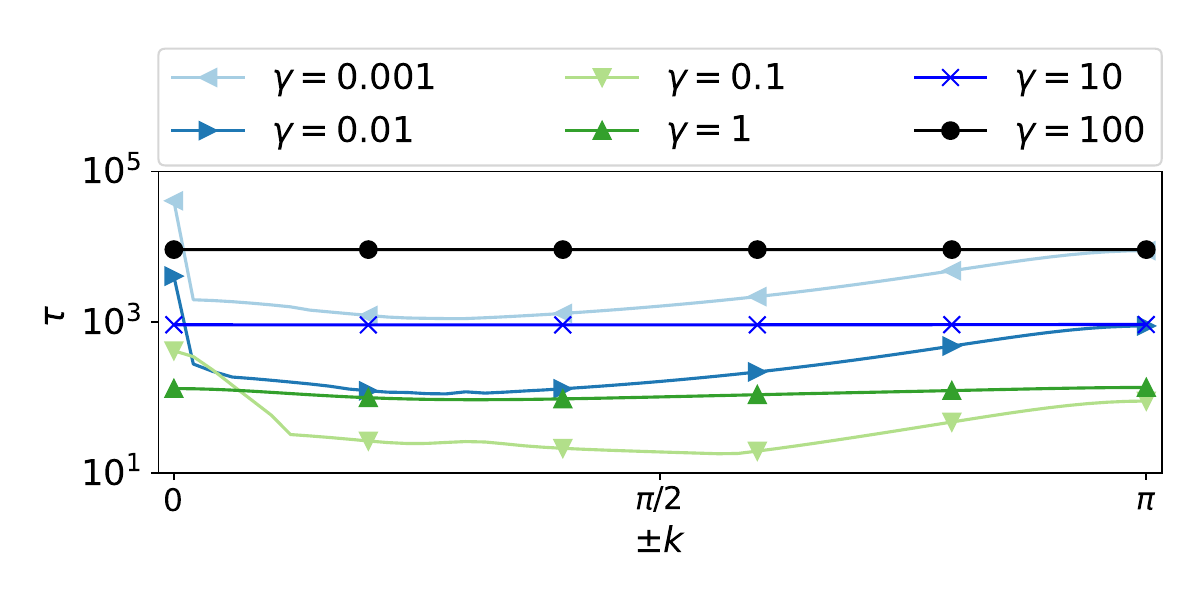}
    \caption{The dependence of $\tau$, as defined in Eq.~\eqref{eq:def_spectral_gap}, on $k$ is shown for various values of  $\gamma$. The  parameters are $L_{n_{-k}}^{n_{+k}}$, $L = 10$, and $J_x = J_y = J_z = 1/3$. The QZE always occurs for $\tau$.
 \label{fig:open_kitaev_tau_B_10_phase_gamma}}
\end{figure}

The ratio $J_\alpha/\gamma$ significantly influences $\tau$.
There is a QZE in $\tau$ for all analyzed jump operators and
Hamiltonian parameters, with a typical example shown in
Fig.~\ref{fig:open_kitaev_tau_B_10_phase_gamma}. The robustness
reaches a minimum at $\gamma \sim J_\alpha$ for $J_x=J_y=J_z$,
consistent with the results for all other analyzed quantities.

\begin{figure}
  \includegraphics[width=\columnwidth]{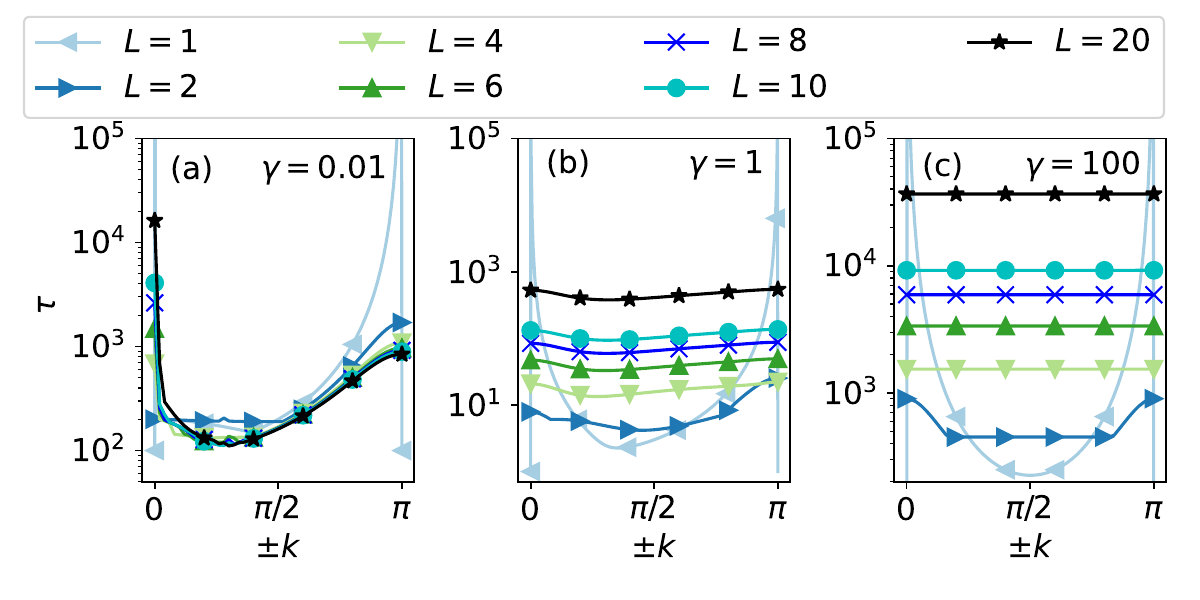}
    \caption{Finite-size effects for $\tau$, as defined in Eq.~\eqref{eq:def_spectral_gap}, are shown. The parameters are $L_{n_{-k}}^{n_{+k}}$ and $J_x = J_y = J_z = 1/3$, and the value of $\gamma$ is indicated in each figure.
 \label{fig:open_kitaev_tau_B_10_phase_parameter}}
\end{figure}

The variation of $\tau$ with $k$ is governed by both $\gamma$ and the
choice of jump operators. The value of $k$ at which $\tau(k)$ attains
its maximum or minimum depends on the parameters of the Hamiltonian
and the jump operators. Both minima and maxima can occur at any value
of $k$. 
While for $\gamma \ll J_\alpha$, $\tau$ can change significantly with
variations in $k$, for $\gamma \gg J_\alpha$, $\tau$ becomes
independent of $k$ for jump operators that induce decoherence, as
shown for one example in
Fig.~\ref{fig:open_kitaev_tau_B_10_phase_gamma}.  
In contrast, for jump operators that do not induce decoherence, $\tau$
exhibits a strong $k$-dependence across all values of
$\gamma$. Additionally, changes in the degeneracy of the eigenvalue
$\lambda_0=0$ (denoted as $\text{deg}(\lambda_0)$), i.e., the number
of steady states, can depend on $k$ for some jump operators and lead
to a divergence of $\tau$ (see
Fig.~\ref{fig:open_kitaev_tau_B_10_phase_parameter}).

$\tau$ can be identical for some physically distinct jump operators but varies strongly for others.
$\tau$ is identical for $L_\mathrm{pc}$ and $L_\mathrm{pa}$, as well as for $L_{n_{+k}}$ and $L_{n_{-k}}$, however, this is not unexpected since the band structures are related through a sign change in these cases. Differences between various sets of jump operators are generally more pronounced at weaker $\gamma$. 
For $\gamma \gg J_\alpha$, $\tau$ becomes identical (and is already nearly identical for $\gamma \approx J_\alpha$) for several combinations (e.g., for $L_{n_{-k}}^{n_{+k}}, L_n$, and $L_\mathrm{pc}^\mathrm{pa}$, or, another example, $L_\mathrm{pc}$, $L_\mathrm{pa}$, $L_{n_{+k}}$, and $L_{n_{-k}}$). Notably, $L_\mathrm{p}$ remains distinct due to differences in $\text{deg}(\lambda_0)$. 
For $\tau$, significant differences emerge between jump operators that preserve and those that break the BdG-type PHS.

Finite-size effects,  as shown in Fig.~\ref{fig:open_kitaev_tau_B_10_phase_parameter}, are weak for  $\gamma \ll J_\alpha$ (except for $k \approx 0$), but $\tau$ can vary by several orders of magnitude for $\gamma \gg  J_\alpha$. 
While for $\gamma \ll J_\alpha$ increasing the system size $L$ slightly decreases $\tau$ (except for $k \approx 0$), for $\gamma \gg J_\alpha$ increasing $L$ strongly increases $\tau$. An exception occurs for $L=1$, where $\tau$ deviates from the trend due to a $k$-dependent $\text{deg}(\lambda_0)$, unlike for $L \geq 2$.

\subsection{Variation of the coupling constants}\label{sec:vary_J_i}

We previously focused mostly on the case $J_x = J_y = J_z$ and now
briefly examine how the results change when the couplings are not
identical.

For unequal coupling parameters $J_x$, $J_y$, and $J_z$, the entropy
exhibits a complex and largely unpredictable time evolution. This
arises from the fact that changing a single parameter can have varying
effects on the entropy, depending on the values of other
parameters. Such behavior is observed across all parameters, including
$J_\mathrm{\alpha}$, $k$, $n_\mathrm{band}$, $L$,  and the choice of
jump operators. 
In particular, the entropy can exhibit fundamentally different
behavior for varying values of $\gamma$. 
As a result, varying a single parameter can lead to either nearly
identical or fundamentally different entropy time evolution.  

Several conclusions follow:  There is no clear relation between the
energy gap of the closed system and the approach to the steady state
in the open system.  The presence or absence of MZMs has no relation
to  the rate of entropy growth. No systematic difference is found in
the entropy increase rate between the KHM and the KC. Different
boundary conditions (OBC or PBC) for the KC can lead to both similar
and distinct entropy dynamics. Entropy time evolution shows no
qualitative distinction between MZMs and other energy bands.

Everything discussed above regarding the entropy also applies to
$\tau$.  However, one clear pattern is observed for $\tau$, which
increases as $J_y$ decreases and ultimately diverges as $J_y$
approaches zero, due to a change in $\text{deg}(\lambda_0)$ (see
Tab.~\ref{tab:KHM_list_steady_state}).  The only effect that
consistently appears for both the entropy and $\tau$ is the QZE. 

Lastly, we examine the time evolution of the band structure under
varying coupling constants. 
Small changes in the couplings $J_\mathrm{\alpha}$ can give rise to
distinct time evolutions, such as strong variations in the timescales
over which energy gaps close. 
Quantitatively (e.g., see the discussion on comparisons between
various jump operators in Sec.~\ref{sec:E_K_time_evol}), the same
effects occur as for identical couplings, but the actual values of
the energies depend strongly on the specific values of the
$J_\mathrm{\alpha}$. 
For example, the rate at which the energies decrease toward zero in
cases where $\rho_\mathrm{mm}$ is the steady state depends strongly on
the couplings $J_\mathrm{\alpha}$ but consistently exhibits the QZE. 
However, not all features of the band structure show the QZE, for
instance, the energy gap in the $A_x$ phase exhibits QZE, whereas in
the $A_y$ and $A_z$ phases, it does not. 
Despite identical band structures in the closed system, the time
evolution differs significantly between the $A_x$ and $A_z$
phases. For example, the energy gap in $A_x$ remains open for several
orders of magnitude longer. 
Finite-size effects, which strongly depend on $\gamma$,
$J_\mathrm{\alpha}$, and the jump operator, influence the energy gap
lifetime, which decreases with increasing system size $L$, approaching
a nonzero value.

\section{Discussion and Conclusions}\label{sec:discuss}

We have investigated the flux-free sector of the KHM under
environmental coupling. This was motivated by the prediction that
signatures of MZMs can be measured with an
STM~\cite{2501.05608,PhysRevB.102.134423,PhysRevLett.126.127201,PhysRevLett.125.267206,2411.01784}. However,
in an STM setup, there is an unavoidable interaction between the
surface and the KHM, which requires analyzing the KHM under
environmental coupling. To do this, we have utilized the LME, focusing
on jump operators that lead to decoherence, to analyze the time
evolution of the entropy and band structure, as well as to calculate
the spectral gap. 

In addition to the expected decoherence induced by number operators,
jump operators based on pair creation/annihilation or incoherent
hopping also cause decoherence, so that the steady state is in most
cases the  maximally mixed state. Exceptions occur for $J_y =0$ or $L = 1$, which correspond to KCs with PBC. Further, jump operators consisting purely of pair creation or pair
annihilation as well as specific incoherent hoppings produce steady
states that are not maximally mixed, but have non-vanishing
off-diagonal elements in the density matrix. 

Variations in the couplings $J_\mathrm{\alpha}$ can lead to
significant differences in $\tau$ and in the time evolution of the
band structure, the entropy, and the fidelity. However, in most cases,
the steady state does not depend on the couplings $J_\mathrm{\alpha}$,
and thus the strong differences only occur on short time
scales. No simple patterns emerge that would allow us to predict whether a variation in $J_\mathrm{\alpha}$ leads to
a faster or slower approach to the steady state.  
However, some statements can be made, e.g., the energy gap closes
significantly faster in the $A_z$ phase compared to the $A_x$ phase. 

The KHM Hamiltonian we investigated reduces to that of a KC for
certain parameter choices, with PBC or OBC depending on the parameters
selected. These parameter sets are among those that have an impact on
time evolution, but one cannot say that the KHM is generally more or
less robust than the KC. 
However, the KC with PBC is a special case in which decoherence does not lead to a maximally mixed steady state.

Under variation of $\gamma$ in the LME, a QZE usually occurs with a
minimum in robustness at $J \sim \gamma$ for $J = J_x = J_y = J_z$. 
The QZE always occurs for the entropy, fidelity, and $\rho(t)$, but
not for all features of the band structure: in the gapped spin liquids
$A_y$ and $A_z$, there is no QZE in the lifetimes of the energy gaps. 

As the system evolves toward a maximally mixed steady state, the
energies of all bands converge to zero. This causes the energy gap to
close and the topological features of the band structure to vanish,
eliminating any distinction between the MZM and the other bands. 
This indicates that a topological phase transition occurs over
time. However, precisely determining this transition requires the
evaluation of a topological invariant, which is beyond the scope of
this work but is an interesting topic for future research. 

Proposals for detecting the topological signatures of the KHM via
STM~\cite{2501.05608,PhysRevB.102.134423,PhysRevLett.126.127201,PhysRevLett.125.267206,2411.01784}
fundamentally rely on the coexistence of a bulk energy gap and gapless
edge states. As the energy gap closes over time, the substrate’s
coupling to the Kitaev monolayer must remain sufficiently weak to
prevent rapid relaxation into the maximally mixed steady state,
thereby preserving the conditions necessary for STM-based observation
of topological signatures. 

While this work focuses on zigzag edges, armchair
edges~\cite{PhysRevB.89.235434, PhysRevB.99.184418} represent an
alternative, and it would be valuable to examine how edge geometry
affects the time evolution. 
Future studies could also use the time evolution of the density matrix
to calculate observables such as spin expectation values, spin-spin
correlations, or the plaquette operator. 
As pure KHM systems are unlikely to be realized
experimentally~\cite{Mandal_2025, 2501.05608}, it is of interest to
extend the model by incorporating additional interactions, for example
by including Heisenberg couplings.

\begin{acknowledgments}
The authors acknowledge support by the state of Baden-Württemberg through bwHPC.
\end{acknowledgments}

\section{References}

\bibliography{bib_file}

\appendix

\section{Numerics}\label{appendix:LME}
One numerical approach to solve the LME, see Eq.~\eqref{eq:lindblad_LME}, is to reformulate it as an eigenvalue problem by utilizing the  Choi–Jamiolkowski isomorphism~\cite{CHOI1975285,JAMIOLKOWSKI1972275}.
This technique, also known as vectorization, involves reshaping the density matrix  into a vector by stacking its columns.
The vectorized form of the LME~\cite{10.1088/1751-8121/aae4d1,PhysRevB.99.174303,1510.08634}  is given by
\begin{equation}
\begin{aligned}
&\frac{\mathrm{d}}{\mathrm{dt}}  \vert\rho\rangle\!\rangle =    \mathcal{L}  \vert\rho\rangle\!\rangle = \biggl[ - i \left( \mathbb{1} \otimes H - H^\mathrm{T} \otimes \mathbb{1} \right)\\
 &+  \gamma \sum_i  \left( {\left(L_i^\dagger \right)}^\mathrm{T} \otimes L_i^{\vphantom{\dagger}} - \frac{1}{2} \left(  \mathbb{1} \otimes  L_i^\dagger  L_i^{\vphantom{\dagger}} +   {\left(L_i^\dagger  L_i^{\vphantom{\dagger}} \right)}^\mathrm{T} \otimes \mathbb{1} \right) \right)  \biggr]  \vert\rho\rangle\!\rangle,
\end{aligned}
\label{eq:linbdlad_vec_LME}
\end{equation}
where   $H$ is the $n \times n $ Hamiltonian of the system,  $\mathbb{1}$ denotes $n \times n $ identity matrix, $\mathcal{L}$ is the  $n^2 \times n^2 $ Lindblad matrix,   and $\vert\rho\rangle\!\rangle$ represents the vectorized density matrix. 
The solution to this differential equation is given by
\begin{subequations} 
\label{eq:LME_vec_solution} 
\begin{align}
 \vert\rho(t)\rangle\!\rangle =  \mathrm{e}^{\mathcal{L}t} \vert\rho(0)\rangle\!\rangle &=  S \mathrm{e}^{\Lambda t} S^{-1}  \vert\rho(0)\rangle\!\rangle \label{eq:LME_vec_solution_exact}\\ 
			& = \left[ \sum_{m=0}^{\infty} \frac{1}{m!} \mathcal{L}^mt^m\right]  \vert\rho(0)\rangle\!\rangle, \label{eq:LME_vec_solution_power_series}
\end{align} 
\end{subequations} 
where the initial density matrix at $t=0$ corresponds  to an eigenstate of the KHM.
For small systems, we can perform full diagonalization (FD) of $\mathcal{L}$ using Eq.~\eqref{eq:LME_vec_solution_exact}, where $S$ is the matrix of eigenvectors and $\Lambda$ is a diagonal matrix comprising the eigenvalues of $\mathcal{L}$.

However,  FD is computationally demanding. 
As an alternative, Eq.~\eqref{eq:LME_vec_solution_power_series} can be employed to obtain solutions for larger systems~\cite{2505.22420}. This approach involves choosing a time step $\Delta t$ and evaluating the power series expansion at $t = 0$ to compute $\vert \rho(\Delta t) \rangle\!\rangle$. The procedure is then iterated to successively determine $\vert \rho(2\Delta t) \rangle\!\rangle$, $\vert \rho(3\Delta t) \rangle\!\rangle$, and so forth.
A suitable choice of parameters is a series expansion truncated at order $m_{\mathrm{max}} = 10$ with a time step $\Delta t = 0.1$. While in some cases lower orders (e.g., $m_{\mathrm{max}} = 5$) and larger time steps (e.g., $\Delta t = 0.5$) may suffice, these parameters are not generally reliable.

Since the dimension of the matrix $\mathcal{L}$ scales as the square of the Hamiltonian's dimension (see Eq.~\eqref{eq:LME_vec_solution}), diagonalizing the Hamiltonian is significantly less computationally demanding than computing $\vert\rho(t)\rangle\!\rangle$. 
Consequently, utilizing  FD for the Hamiltonian is sufficient.

\section{Jump operators}\label{sec_appendix_jump_operators}
Table~\ref{tab:KHM_list_jump_operators} contains the definitions of the jump operators examined and indicates whether these operators lead to decoherence and whether they break or preserve the PHS of the BdG formalism.
Table~\ref{tab:KHM_list_steady_state} lists the steady states and their degeneracies.

\renewcommand{\arraystretch}{1.3}
\begin{table}[t]
\centering
   \caption{ Definitions of the jump operator sets used in the LME are summarized. The leftmost column lists the labels for each set, while the adjacent column specifies the corresponding jump operators. Each set is constructed as a sum over the site index $l$, with uniform dissipation strength $\gamma$ applied throughout. The third column indicates whether the jump operators induce decoherence, when this depends on Hamiltonian parameters, the specific conditions are provided. Here, $n \in \mathbb{Z}$. The rightmost column identifies whether the set breaks the BdG-type PHS. The notation used includes "pc" and "pa" for pair creation and pair annihilation operators, respectively, and "R" and "L" to denote hopping to the right (from the site $l+1$ to $l$) and to the left (from the site $l$ to $l+1$), respectively.  }
    \begin{tabular}{l  l l l }
    \hline \hline   
label &  jump operator & decoherence & BdG-type PHS \\\hline
$L_{n_{+k}}$ & $ L_l= \alpha_{(l,k)}^\dagger   \alpha_{(l,k)}^{\vphantom{\dagger}}$ & yes &  broken if: $L\geq 2$  \\
&&& preserved if: $L=1$ \\
 $L_{n_{-k}}$ &    $L_l=\alpha_{(l,-k)}^{\vphantom{\dagger}}   \alpha_{(l,-k)}^\dagger $  & yes  & broken if:  $L\geq 2$:   \\
&&& preserved if: $L=1$ \\
 $L_n$  & $L_l=\alpha_{(l,k)}^\dagger   \alpha_{(l,k)}^{\vphantom{\dagger}} -  \alpha_{(l,-k)}^{\vphantom{\dagger}}  \alpha_{(l,-k)}^\dagger   $ & yes &  preserved  \\
$L_{n_{-k}} ^{n_{+k}} $   &  $L_l^+ =\alpha_{(l,k)}^\dagger   \alpha_{(l,k)}^{\vphantom{\dagger}}$  & yes &  preserved  \\
& $L_l^- = \alpha_{(l,-k)}^{\vphantom{\dagger}}  \alpha_{(l,-k)}^\dagger   $    & &  \\
$L_{\mathrm{pc}}$  & $ L_l= \alpha_{(l,k)}^\dagger   \alpha_{(l,-k)}^\dagger$  & $L=1$, $J_z=0$: yes & broken   \\
&&  $L=1$,  $J_z\neq 0$, $k = n \pi$: yes & \\
&&else: no& \\
$L_{\mathrm{pa}}$   & $ L_l= \alpha_{(l,-k)}^{\vphantom{\dagger}}   \alpha_{(l,k)}^{\vphantom{\dagger}}$ & $L=1$,  $J_z=0$ : yes &  broken  \\
&&  $L=1$,  $J_z\neq 0$, $k = n \pi$: yes & \\
&&else: no& \\
$L_{\mathrm{p}}$   & $ L_l= \alpha_{(l,k)}^\dagger   \alpha_{(l,-k)}^\dagger + \alpha_{(l,-k)}^{\vphantom{\dagger}}   \alpha_{(l,k)}^{\vphantom{\dagger}}$   & $J_x=J_z$, $k=2\pi n$: no & preserved  \\
& & else: yes & \\
$L_{\mathrm{pc}}^{\mathrm{pa}}$   & $ L_l^\mathrm{c}= \alpha_{(l,k)}^\dagger   \alpha_{(l,-k)}^\dagger$    & yes & preserved \\
& $L_l^\mathrm{a}= \alpha_{(l,-k)}^{\vphantom{\dagger}}   \alpha_{(l,k)}^{\vphantom{\dagger}}$  & &  \\
$L_\mathrm{J,R}^{+k}$  & $L_l = \alpha_{(l,k)}^\dagger   \alpha_{(l+1,k)}^{\vphantom{\dagger}}$ & no  & broken  \\
 $L_\mathrm{J,L}^{+k}$ & $L_l = \alpha_{(l+1,k)}^\dagger   \alpha_{(l,k)}^{\vphantom{\dagger}}$ & no  &broken\\
$L_\mathrm{J,L}^{-k}$ & $L_l= \alpha_{(l,-k)}^{\vphantom{\dagger}}   \alpha_{(l+1,-k)}^\dagger$ & no   &broken  \\
$L_\mathrm{J,R}^{-k}$& $L_l = \alpha_{(l+1,-k)}^{\vphantom{\dagger}}   \alpha_{(l,-k)}^\dagger$ & no  &broken  \\
$L_\mathrm{J}$  & $L_l^{(1)} = \alpha_{(l,k)}^\dagger   \alpha_{(l+1,k)}^{\vphantom{\dagger}}$ &  yes  & preserved  \\
 & $L_l^{(2)} = \alpha_{(l+1,k)}^\dagger   \alpha_{(l,k)}^{\vphantom{\dagger}}$ &  &  \\
& $L_l^{(3)} = \alpha_{(l,-k)}^{\vphantom{\dagger}}   \alpha_{(l+1,-k)}^\dagger$ &   &  \\
& $L_l^{(4)} = \alpha_{(l+1,-k)}^{\vphantom{\dagger}}   \alpha_{(l,-k)}^\dagger$ &   &  \\
$L_\mathrm{J}^{+k}$  & $L_l^{(1)} = \alpha_{(l,k)}^\dagger   \alpha_{(l+1,k)}^{\vphantom{\dagger}}$ & yes  & preserved  \\
 & $L_l^{(2)} = \alpha_{(l+1,k)}^\dagger   \alpha_{(l,k)}^{\vphantom{\dagger}}$ &&    \\
$L_\mathrm{J}^{-k}$& $L_l^{(1)} = \alpha_{(l,-k)}   \alpha_{(l+1,-k)}^\dagger$ &yes   & preserved  \\
& $L_l^{(2)} = \alpha_{(l+1,-k)}^{\vphantom{\dagger}}   \alpha_{(l,-k)}^\dagger$ & &    \\
$L_\mathrm{J,R}$  & $L_l^{(1)} = \alpha_{(l,k)}^\dagger   \alpha_{(l+1,k)}^{\vphantom{\dagger}}$ & no  & preserved  \\
 & $L_l^{(2)}= \alpha_{(l+1,-k)}^{\vphantom{\dagger}}   \alpha_{(l,-k)}^\dagger$ &&     \\
$L_\mathrm{J,L}$  & $L_l^{(1)} = \alpha_{(l+1,k)}^\dagger   \alpha_{(l,k)}^{\vphantom{\dagger}}$ & no  & preserved  \\
 & $L_l^{(2)}= \alpha_{(l,-k)}^{\vphantom{\dagger}}   \alpha_{(l+1,-k)}^\dagger$ & &   \\
\hline  \hline
\end{tabular}
\label{tab:KHM_list_jump_operators}
\end{table}
\renewcommand{\arraystretch}{1}

\renewcommand{\arraystretch}{1.3}
\begin{table}[t]
\centering
   \caption{  Steady-state density matrices corresponding to the sets of jump operators defined in Tab.~\ref{tab:KHM_list_jump_operators} are summarized. The steady state is the right eigenvector of $\mathcal{L}$ (see Eq.~\eqref{eq:linbdlad_vec_LME}) associated with the eigenvalue $\lambda_0 = 0$. When the steady state depends on Hamiltonian parameters, these conditions are specified in the "conditions" column. The variable $n \in \mathbb{Z}$. Here, $\text{diag}(\dots)$ indicates a diagonal matrix, and $\deg(\lambda_0)$ denotes the degeneracy of the  eigenvalue $\lambda_0$. The maximally mixed state, as defined in Eq.~\eqref{eq:kitaev_max_mixed_state}, is represented by $\rho_{\mathrm{mm}}$. } 
    \begin{tabular}{l  l l l}
    \hline \hline   
jump operators &  conditions & deg$(\lambda_0)$ & steady state \\\hline
$L_{n_{+k}}$,  $L_{n_{-k}}$,  $L_n$ , $L_{n_{-k}} ^{n_{+k}} $    & $ L\geq 2$,  $J_y \neq 0$  & $1$ & $\rho_\mathrm{mm}$  \\
& $ L\geq 2$,  $J_y = 0$  & $\neq 1$ & diagonal   \\
& $L=1$, $J_z\neq0$, $k \neq n \pi $ & 1 &  $\rho_\mathrm{mm}$  \\
& $L=1$, $J_z\neq0$, $k = n \pi $ & 2 & diag$(1,0)$, diag$(0,1)$  \\
& $L=1$, $J_z=0$ & 2 & diag$(1,0)$, diag$(0,1)$ \\
$L_{\mathrm{pc}}$  &$L=1$,  $J_z=0$ &1  & diag$(1,0)$ \\
&$L=1$,  $J_z \neq 0$, $k= n \pi$ &1  & diag$(1,0)$ \\
& $L \geq 2$, $J_y =0$ & $\neq 1$ & non-diagonal \\
& else & $1$ & non-diagonal  \\
$L_{\mathrm{pa}}$  &$L=1$,  $J_z=0$ &1  & diag$(0,1)$ \\
&$L=1$,  $J_z \neq 0$, $k= n \pi$ &1  & diag$(0,1)$ \\
& $L \geq 2$, $J_y =0$ & $\neq 1$ & non-diagonal \\
& else & $1$ & non-diagonal  \\
$L_{\mathrm{p}}$ & $J_y \neq 0$, $J_x=J_z$,  $k=2\pi n$ & $ \neq 1$ & non-diagonal \\
& $L\geq 2$, $J_y =0$ & $\neq 1$  &  non-diagonal  \\
& else&$1$ &  $\rho_\mathrm{mm}$ \\
$L_{\mathrm{pc}}^{\mathrm{pa}}$    & $L \geq 2$,  $J_y =0$ & $\neq 1$ & non-diagonal  \\
& $L \geq 2$,  $J_y \neq0$  & $1$ &  $\rho_\mathrm{mm}$ \\
& $L=1$  & $1$ &  $\rho_\mathrm{mm}$ \\
$L_\mathrm{J,R}^{+k}$,  $L_\mathrm{J,L}^{+k}$, $L_\mathrm{J,L}^{-k}$, $L_\mathrm{J,R}^{-k}$ & $J_y \neq 0$ & 1 & non-diagonal \\ 
& $J_y = 0$ & $\neq 1$ & non-diagonal  \\
$L_\mathrm{J,R}$, $L_\mathrm{J,L}$ & $J_y \neq 0$ & 1 & non-diagonal \\ 
& $J_y = 0$ & $\neq 1$ & non-diagonal  \\
$L_\mathrm{J}$,  $L_\mathrm{J}^{+k}$,  $L_\mathrm{J}^{-k}$ & $J_y \neq 0$ & 1 & $\rho_\mathrm{mm}$ \\
& $J_y = 0$ & $\neq 1$ & non-diagonal  \\
\hline \hline
\end{tabular}
\label{tab:KHM_list_steady_state}
\end{table}
\renewcommand{\arraystretch}{1}

\end{document}